\documentclass[twocolumn,,amsmath,amssymb]{revtex4-1}

\usepackage{graphicx}
\usepackage{dcolumn}
\usepackage{bm}
\newcommand{\etal}{\emph{et al.}}
\newcommand{\be}{\begin{equation}}
\newcommand{\ee}{\end{equation}}
\newcommand{\bfig}{\begin{figure}}
\newcommand{\efig}{\end{figure}}

\usepackage{lineno}
\usepackage{lipsum}

\begin{document} 

\title{The planar thermal Hall conductivity in the Kitaev magnet $\alpha$-RuCl$_3$.
}

\author{Peter Czajka$^{1}$}
\author{Tong Gao$^{1}$}
\author{Max Hirschberger$^{2}$}
\author{Paula Lampen-Kelley$^{3,4}$}
\author{Arnab Banerjee$^{6}$}
\author{Nicholas Quirk$^{1}$}
\author{David G. Mandrus$^{3,4}$}
\author{Stephen E. Nagler$^{5}$}
\author{N. P. Ong$^{1,\S}$}
\affiliation{
{$^1$Department of Physics, Princeton University, Princeton, NJ 08544, USA}\\
{$^2$Dept. of Applied Physics, The University of Tokyo, 7-3-1 Hongo, Bunkyo-ku, Tokyo 113-8656, Japan}\\
{$^3$Department of Materials Science and Engineering, University of Tennessee, Knoxville, Tennessee 37996, USA} \\
{$^4$Materials Science and Technology Division, and $^5$Neutron Scattering Division, Oak Ridge National Laboratory, Oak Ridge, Tennessee 37831, USA}\\
{$^6$Department of Physics and Astronomy, Purdue University, West Lafayette, IN 47907-2036}
}

\date{\today} 

\begin{abstract}
We report detailed measurements of the Onsager-like planar thermal Hall conductivity $\kappa_{xy}$ in $\alpha$-RuCl$_3$, a spin-liquid candidate of topical interest. With the thermal current ${\bf J}_{\rm Q}$ and magnetic field $\bf B\parallel a$ (zigzag axis), the observed $\kappa_{xy}/T$ varies strongly with temperature $T$ (1-10 K). The results are well-described by bosonic edge excitations which evolve to topological magnons at large $B$. Fits to $\kappa_{xy}/T$ yield a Chern number $\sim 1$ and a band energy $\omega_1\sim$1 meV, in agreement with sharp modes seen in electron spin-resonance experiments. The bosonic character is incompatible with half-quantization of $\kappa_{xy}/T$.

\end{abstract}
\maketitle 


In the field of quantum spin liquids~\cite{Savary,Ng,Takagi}, the honeycomb magnet $\alpha$-RuCl$_3$ has attracted considerable interest because it is proximate to the Kitaev Hamiltonian ${\cal H}_K$~\cite{Khaliullin,Plumb,Banerjee,Sears2,Loidl,Leahy,Banerjee2,Hentrich,Lampen}. In 2006, Kitaev~\cite{Kitaev} published the exact solution for the ground state of ${\cal H}_K$, and found that the excitations are Majoranas and vortices. The Majorana excitations lead to a thermal Hall conductivity $\kappa_{xy}$ that is half-quantized.

The thermal Hall effect provides a powerful probe of spin excitations, for e.g. in the pyrochlore Tb$_2$Ti$_2$O$_7$~\cite{Hirschberger} and the kagome magnet Cu(1-3, bdc)~\cite{YoungLee}. In 2018, evidence for a half-quantized $\kappa_{xy}$ in $\alpha$-RuCl$_3$ in the narrow interval 3.7 $<T<$ 5 K was reported in Ref. \cite{Kasahara}. Half-quantization of $\kappa_{xy}/T$ was also observed~\cite{Yokoi} with $\bf B\parallel a$ but in a narrow field interval distinct from that in Ref. \cite{Kasahara}. Subsequently, the authors of Ref. \cite{Czajka} reported that the planar $\kappa_{xy}$ was strongly $T$ dependent showing no trace of half-quantized behavior. In this report, large oscillations of $\kappa_{xx}$ were observed with maximal amplitude within the window $7<H<11.5$ T. A third group recently reported observing a robust half-quantization signal in the planar geometry~\cite{Takagi21}. The conflicting results reflect the weak signal and large hysteretic and magnetocaloric effects below 4 K.

Adopting measures to mitigate these distortions, we have obtained $\kappa_{xy}$ over a broad interval of $T$ (0.5 to 10 K). Our results point to a view of $\kappa_{xy}$ categorically distinct from the half-quantized picture. We show that $\kappa_{xy}$ originates from bosonic edge-mode excitations that become topological magnons at large $B$. Whereas fermionic edge modes have been investigated intensively in quantum Hall systems, our experiment is possibly the first demonstration of neutral bosonic edge modes in a magnetic insulator.

\begin{figure*}[t]
\includegraphics[width=18 cm]{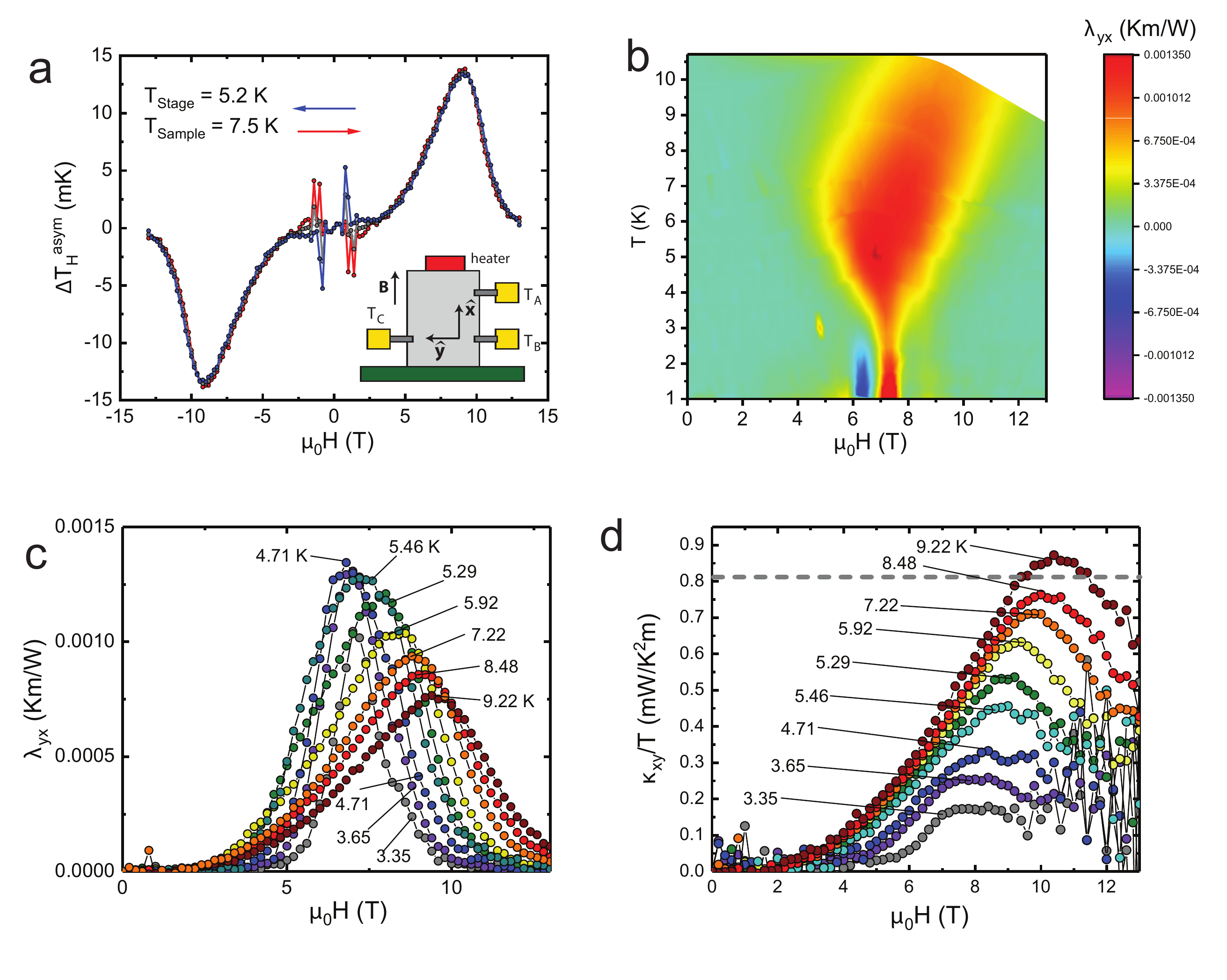}
\caption{\label{figKxyColor} 
Planar thermal Hall response of $\alpha$-RuCl$_3$ in the interval 0.5 $< T<$ 10.5 K and field range $0<B< 13$ T with $\bf B\parallel a$. Panel (a) shows the $B$-antisymmetric thermal Hall signal $\Delta T_{\rm H}^{\rm asym}$ taken in sweep-up (red) and -down (blue) directions at 7.5 K ($\Delta T_{\rm H}^{\rm asym}$ is proportional to $\lambda_{yx}$). The inset shows the placements of thermometers measuring $T_{\rm A}$, $T_{\rm B}$ and $T_{\rm C}$. Panel (b): The color map constructed from $\sim 50$ traces of the thermal Hall resistivity $\lambda_{yx}$ vs. $B$ in the $B$-$T$ plane (scale bar on right). The large-$\lambda_{yx}$ region (in red) tapers down to a neck around 7.3 T, as $T\to 0.4$ K. Below 2.5 K, $\lambda_{yx}<0$ in a sliver at 6.3 T (blue region). Panel (c) shows curves of $\lambda_{yx}$ vs. $B$ at fixed $T$. The peak in the field profiles narrows as $T\to$ 3.35 K. Below 4 K, $\lambda_{yx}$ is strongly suppressed to zero above 9.5 T. Panel (d) plots $\kappa_{xy}/T$ vs. $B$ (at fixed $T$) inferred from $\lambda_{ij}$. The non-intersecting curves imply that, at any fixed $B$, $\kappa_{xy}/T$ is monotonic in $T$. Below 4.7 K the matrix inversion greatly amplifies the uncertainties in $\lambda_{xy}$ for $B>$9 T. 
}
\end{figure*}

In the experiment, both the thermal current ${\bf J}_{\rm Q}$ and $\bf B$ are applied $\parallel {\bf \hat{x}\parallel a}$. During each run lasting 8-10 h, $B$ is incremented in steps $\delta B =$ 0.2 T from -13 to +13 T (and back to -13 T) while $T$ is regulated within $\pm 1$ mK of its set-point. We adopt a step-probe protocol in which measurements of the sensors $T_{\rm A}$, $T_{\rm B}$ and $T_{\rm C}$ are delayed by 150 s after each step-change. The protocol rigorously excludes contamination by the large transients caused by magnetocaloric and eddy-current heating effects. Hysteretic artefacts are eliminated by combining field sweep-up and -down curves. The 3 readings determine the values of $\lambda_{yx}$ and $\lambda_{xx}$ at each $B$.

Figure \ref{figKxyColor} (a) shows the explicitly $B$-antisymmetric thermal Hall signal $\Delta T_{\rm H}^{\rm asym}$ (proportional to the thermal Hall resistivity $\lambda_{yx}$) measured at 7.5 K in an in-plane $\bf B\parallel \hat{x}$ (inset). At each $B$, the thermal Hall signal is obtained by combining the two readings $T_{\rm B}-T_{\rm C}$ and $T_{\rm A}-T_{\rm C}$, and field-antisymmetrizing. In Panel (b), the color map provides an overview of how $\lambda_{yx}$ varies over the $B$-$T$ plane. The large-$\lambda_{yx}$ region above 4 K (red area) tapers to a thin neck at 7.5 T as $T$ decreaes to 1 K. Below 2.5 K, $\lambda_{yx}$ becomes slightly negative in the small region shown in blue. 
The evolution of the field profiles of $\lambda_{yx}$ is shown in Fig. \ref{figKxyColor}c. At 9.22 K, the profile features a broad peak that narrows dramatically as $T\to$ 3.65 K. At high $B$ ($> 10$ T), $\lambda_{yx}$ is strongly suppressed to values below our resolution for $T<$ 4 K.

From the matrix inversion $\kappa_{ij} = [\lambda^{-1}]_{ij}$, we derive the curves of $\kappa_{xy}/T$ vs. $B$ which display a peak in the field profile (Fig. \ref{figKxyColor}d). In the regime $B>$ 10 T and $T<$ 4 K, the matrix inversion leads to large uncertainties (caused by multiplying a near-zero $\lambda_{yx}$ by a factor of $\sim 10^3$). Over the interval $0.5<T<10$ K, $\kappa_{xy}/T$ does not show evidence of half-quantization (dashed line). Instead, the pervasive feature is strongly $T$-dependent $\kappa_{xy}$.

\begin{figure*}[t]
\includegraphics[width=16 cm]{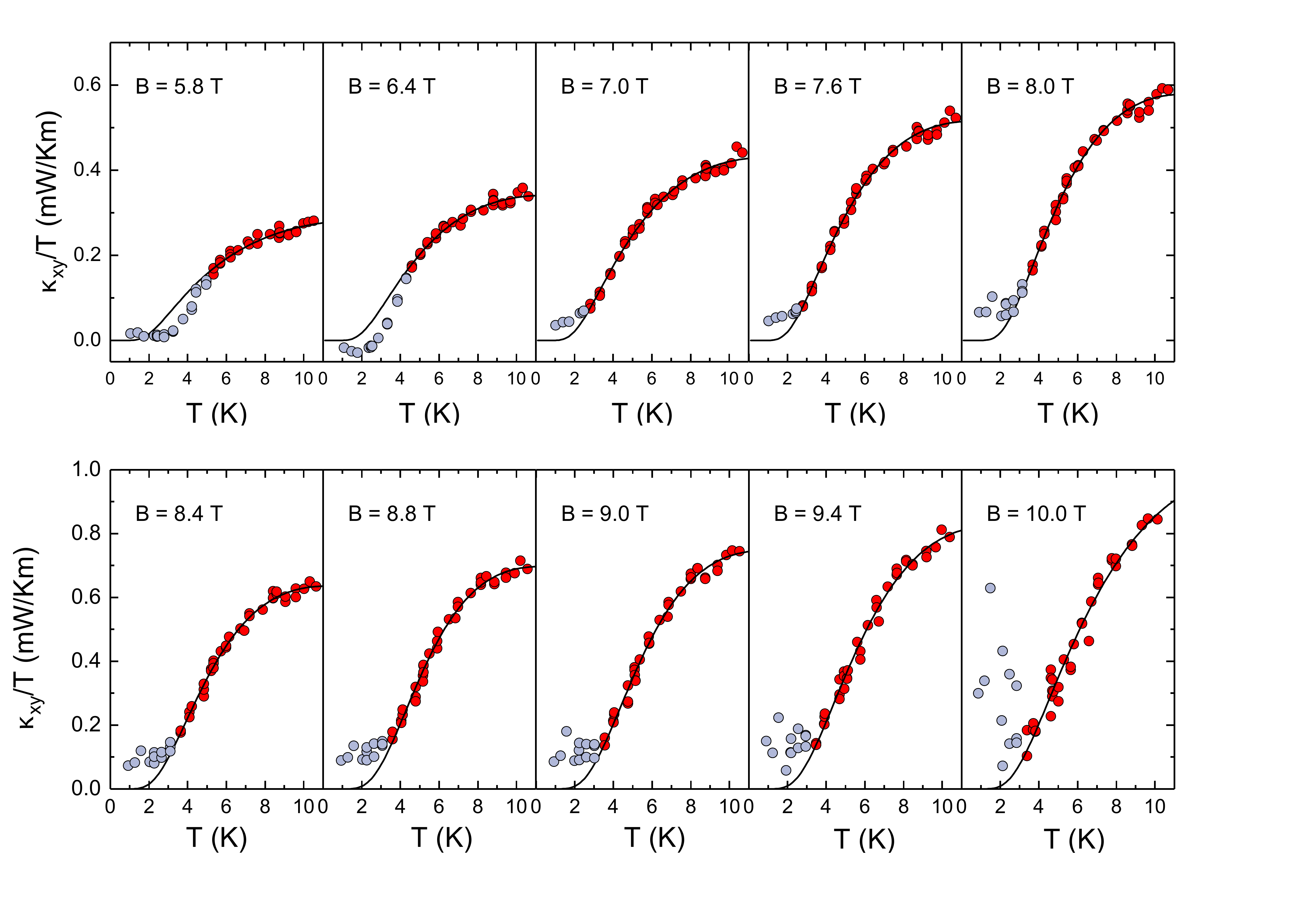}
\caption{\label{figFits} 
Curves of $\kappa_{xy}/T$ vs. $T$ at $B$ fixed at the ten values indicated. In each panel, the solid curves are fits to Eqs. \ref{KH} and \ref{c2}. At each $B$, the fit yields the energies $\omega_1$ and $\omega_2$ and ${\cal C}_{\rm obs}$. The grey circles represent weak deviations from the fits below 4 K. Just below the boundary of the zigzag phase ($B<$7 T), excitations in the zigzag state lead to negative deviations (see also Fig. \ref{figOmega}b). In the QSL state (7$<B<$ 11 T), positive deviations onset below 3 K and grow in amplitude as $T\to 0$. These seem to be related to the excitations that lead to oscillations in $\kappa_{xx}$ (see Supplement).
}
\end{figure*}

Figure \ref{figFits} shows how $\kappa_{xy}/T$ varies with $T$ with $B$ fixed at between 5 and 10 T. At each value of $B$, $\kappa_{xy}/T$ falls monotonically towards zero as $T$ decreases below 10 K. We have found that the steep $T$ dependence is in quantitative agreement with edge modes populated by bosons (solid curves).

First, a planar thermal Hall effect that is odd in $\bf B$ is quite unexpected. In topological materials, however, the emergence of a Berry curvature $\bf \Omega$ that reverses sign with $\bf B$ can lead to such a Hall response, as seen in ZrTe$_5$~\cite{Liang}. For this purpose, we would need $\bf \Omega\parallel \hat{c}^*$ (normal to the honeycomb layer) as well as to reverse sign with $\bf B\parallel a$.

\begin{figure*}[t]
\includegraphics[width=18 cm]{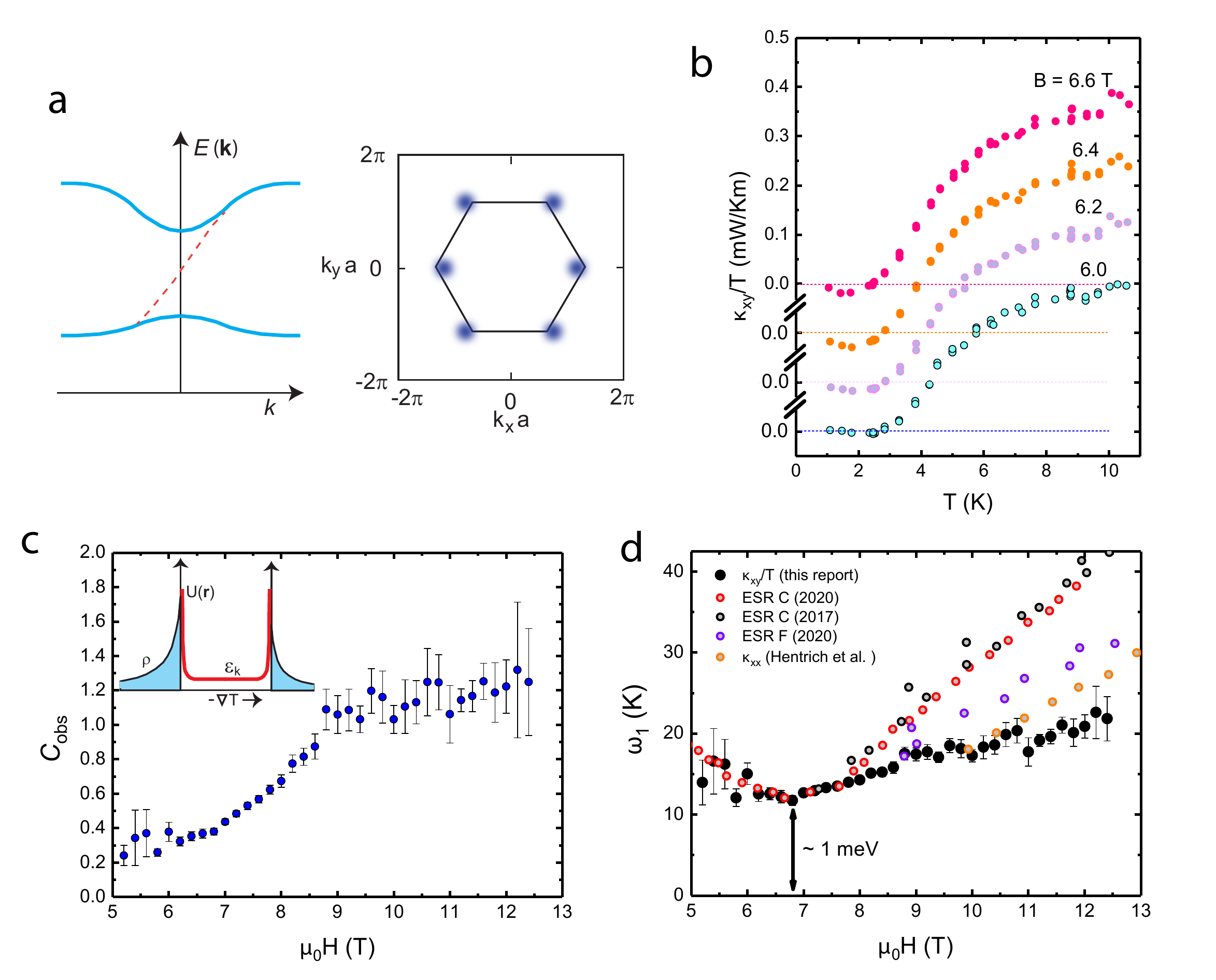}
\caption{\label{figOmega} Bosonic edge-mode and the planar $\kappa_{xy}$.
Panel(a) shows a sketch of the two lowest magnon modes (blue curves) with $\omega_1$ = 11.6 K and $\omega_2\simeq$ 50 K. With ${\cal C} = 1$ in the lowest band, an edge mode traverses the gap (dashed curve). The inset displays the high-intensity spots of the Berry curvature $\bf \Omega$ calculated~\cite{Chern} for the lowest band with $\bf B\parallel a$. $\bf \Omega$ changes sign with $\bf B$.
Panel (b): Curves of $\kappa_{xy}/T$ vs. $T$ with $B$ fixed at 6.0, 6.2, 6.4 and 6.6 T (just inside the zigzag phase). Below 3 K, $\kappa_{xy}$ displays a weak negative contribution. For clarity, we have shifted successive curves vertically by 0.1 mW/Km. 
Panel (c): The quantity $C_{\rm obs}$ defined in Eq. \ref{Cobs} derived from the fit at each $B$. As $B$ increases, $C_{\rm obs}$ saturates to a value $\sim$1.1 above 9 T, consistent with the Chern integer 1. In the inset, the profile of $U({\bf r})+\omega_1$ is sketched as the red curve. A gradient $-\nabla T\parallel\bf\hat{x}$ unbalances the excitation densities between the warm and cool edges to produce ${\bf J}_{\rm Q}\parallel{\bf\hat{y}}$ (distribution $\rho$ shaded in blue). Panel (d) shows the $B$ dependence of the energy level $\omega_1$ derived from the fit of $\kappa_{xy}/T$ vs. $T$ to Eq. \ref{K12} at each $B$. At the minimum (at 6.8 T), $\omega_1$ agrees with the energy of the narrow mode seen in ESR ($\sim$1 meV). At large $B$, $\omega_1$ is slightly lower than the ESR mode, and in better agreement with the energy extracted from $\kappa_{xx}$ in Ref. \cite{Hentrich}.
}
\end{figure*}

Recent calculations have shown that $\bf \Omega$ indeed emerges in $\alpha$-RuCl$_3$ with high-intensity spots near each of the $K$ points in the Brillouin zone (BZ)~\cite{McClarty,Joshi,Chern} (Fig. \ref{figOmega}a, inset). At large $B$, the excitations, called topological magnons, occupy bands in which the Chern number ${\cal C}_n$ alternates in sign~\cite{McClarty,Joshi,Chern}. 
Significantly, when $\bf B\parallel a$, $\bf \Omega$ reverses sign with $\bf B$~\cite{Chern,Zhang}.

In a 2D magnet with finite $\bf\Omega$, the thermal Hall conductivity is given by~\cite{Matsumoto,Murakami}
\be
\frac{\kappa_{xy}}{T} = \frac{1}{\hbar V}\sum_{n,{\bf k}} \Omega_{n,z}({\bf k}) 
			\int_{\varepsilon_{n,{\bf k}}}^\infty d\varepsilon \frac{(\varepsilon-\mu)^2}{T^2}
			\left(-\frac{d\rho}{d\varepsilon}\right),
\label{Kmura}
\ee
where the sum is over bands with dispersion $\varepsilon_{\bf k}$ and Berry curvature ${\bf\Omega}_n$, $\rho$ is the distribution function of the relevant excitaion, $V$ is the sample volume, $\mu$ the chemical potential and $\hbar = h/2\pi$ with $h$ the Planck constant.

Semiclassically, we may regard a wave-packet subject to $\bf\Omega$ and the force $-\nabla U$ exerted by the wall potential $U$~\cite{Murakami} (Fig. \ref{figOmega}a, inset). The anomalous velocity ${\bf v}_A = -\nabla U\times \bf\Omega$ drives a circulating thermal current around the edges. A thermal gradient $-\nabla T\parallel \bf\hat{x}$ unbalances the excitation density between the warm edge $\parallel\bf\hat{y}$ and the cool edge, which leads to a net thermal current ${\bf J}_{\rm Q}\parallel \bf\hat{y}$ (Fig. \ref{figOmega}c). Crucially, the reversal of $\bf \Omega$ induced by reversing $\bf B$ leads to an Onsager planar thermal Hall current.

If the excitations are fermions, Eq. \ref{Kmura} yields the ($T$-independent) Kane-Fisher result~\cite{KaneFisher,Sachdev}
\be
\frac{\kappa_{xy}}{T}= \frac{\pi^2}{3}\frac{k_B^2}{h} \nu, \quad (\nu\in Z).
\label{Kane}
\ee
where $k_B$ is Boltzmann's constant.

By contrast, for bosons, Eq. \ref{Kmura} yields a very strong $T$ dependence. In units of the universal thermal conductance $k_B^2/h$ (and setting $\mu=0$), Eq. \ref{Kmura} simplifies to
\be
\frac{{\cal K}_{\rm H}}{T} \equiv \frac{\kappa_{xy}/T}{k_B^2/h} = \sum_n {\cal C}_n\; c_{2}^{(n)}(\omega_n,T)
\label{KH}
\ee
The function $c_{2}^{(n)}(\omega_n,T)$ is defined as~\cite{Matsumoto,Murakami} 
\be
c_{2}^{(n)}(\omega_n,T) = \int_{u_{0n}}^\infty du\,u^2(-d\rho/du), \quad (u_{0n} = \beta\omega_n({\bf k})),
\label{c2}
\ee
where $\rho = 1/({\rm e}^u -1)$ and $\beta = 1/(k_BT)$. The lowest band is nearly flat~\cite{Chern}. Adopting the flat-band approximation, we used the winding-number equation $2\pi{\cal C}_n = \int_{\rm BZ} d^2k\; \Omega_{n,z}({\bf k})$ to relate $\Omega_{n,z}$ to ${\cal C}_n$.

At low $T$, we retain the 2 lowest bands in Eq. \ref{KH} (Fig. \ref{figOmega}a and Supplement). Assuming that their Chern numbers alternate in sign (${\cal C}_1 = -{\cal C}_2$), we have 
\be
\frac{{\cal K}_{\rm H}}{T} = {\cal C}_1\left[c^{(1)}_2(\omega_1,T)-c^{(2)}_2(\omega_2,T)\right].
\label{K12}
\ee
The overall scale of ${\cal K}_{\rm H}/T$ is fixed by ${\cal C}_1$.

We find that Eq. \ref{K12} provides close fits to the observed $\kappa_{xy}/T$ over a broad range of $B$ (the fits were carried out on $\kappa_{xy}^{\rm 2D} = \kappa_{xy}d$ with $d$ = 5.72 $\AA$). 
In Fig. \ref{figFits}, the fits are shown as the solid curves. We have found that slight deviations from the fits (grey circles) reveal important information on the excitations.

For $B<$ 6.8 T, $\kappa_{xy}/T$ displays a weak, negative deviation (dip) below 3 K (in the blue region in Fig. \ref{figKxyColor}b). 
The expanded view in Fig. \ref{figOmega}b shows the dips in 4 traces of $\kappa_{xy}/T$ vs. $T$. The negative dips imply that, once we enter the ordered phase (below 7 T), the spin excitation branches deviate from Eq. \ref{KH}. The weak negative dip feature also appears in calculations of $\kappa_{xy}$ (see Figs. 3 and 6 in Ref. \cite{Zhang}).
Weak, positive deviations are also observed below $\sim$10 T. These deviations represent the growth of excitations in the QSL state as $T\to 0$. Within the uncertainties, they scale with the amplitude of the oscillations in $\kappa_{xx}$ (Supplement).

Away from these deviations, the fits shown in Fig. \ref{figFits} describe quite accurately the strong $T$ dependence of $\kappa_{xy}/T$ over a broad range of $B$. At each $B$, the fit yields the 2 energies $\omega_1$ and $\omega_2$ (in Eq. \ref{c2}) as well as the dimensionless quantity ${\cal C}_{\rm obs}$ discussed next.

Dividing Eq. \ref{K12} across by the $T$-dependent factor $[c^{(1)}_2(\omega_1,T)-c^{(2)}_2(\omega_2,T)]$, we define 
\be
{\cal C}_{\rm obs} \equiv ({\cal K}_{\rm H}/T)/[c^{(1)}_2(\omega_1,T)-c^{(2)}_2(\omega_2,T)], 
\label{Cobs}
\ee
which can be compared with ${\cal C}_1$. ${\cal C}_{\rm obs}$ is plotted vs. $B$ in Fig. \ref{figOmega}c.

We find that ${\cal C}_{\rm obs}$ starts off small ($\sim 0.3$ at 5 T) but increases to attain a plateau above 9 T. Within the experimental uncertainty, the nearest integer at the plateau is ${\cal C}_{\rm obs}\simeq 1$. 
The strong $T$ dependence of $\kappa_{xy}/T$ largely arises from the bosonic distribution in the integrand of $c^{(n)}_2$ (Eq. \ref{c2}). Remarkably, once this is factored out, the overall amplitude in the polarized state at high fields is fixed by the integer ${\cal C}_{\rm obs}\simeq 1$. 

We show next that the energy $\omega_1$ derived from the fits of $\kappa_{xy}/T$ to Eq. \ref{Kmura} closely agrees with the energy of the dominant sharp mode seen in electron spin resonance (ESR)~\cite{Ponomaryov1,Ponomaryov2}, microwave absorption~\cite{Loidl,Wellm} and neutron-scattering experiments~\cite{Balz}. The energy of the second band is found to be $\omega_2\sim 50 \pm 10$ K with a large uncertainty.
In the field interval above $B_c$ (from 7.3 to 14 T), $\alpha$-RuCl$_3$ displays a rich magnetic resonance spectrum with 4 sharp modes ($C$, $D$, $E$ and $F$)~\cite{Ponomaryov1} superposed on a broad continuum suggestive of excitations~\cite{Loidl,Wellm}. The broad continuum was observed also in microwave absorption and inelastic neutron scattering experiments~\cite{Balz}. At $B$= 7.3, the two lowest modes $C$ and $F$ are degenerate at 0.27 THz (1.1 meV). As $B$ is raised to 16 T, both increase steeply to 1.1 THz (for $C$) and 0.9 THz ($F$). These modes are also seen in exact diagonalization of a 24-spin model~\cite{Winter}. Close to $B_c$, $\omega_1$ inferred from $\kappa_{xy}/T$ agrees remarkably well with the degenerate values of $C$ and $F$. At large $B$, $\omega_1$ is lower than $C$ ($C$ arises from vertical $\Delta{\bf q}=0$ transitions~\cite{Ponomaryov2} at the zone center $\Gamma$). This suggests that, as $C$ rises steeply with $B$, $\kappa_{xy}$ is weighted towards lower-lying excitations from elsewhere in the BZ. The curve of $\omega_1(B)$ agrees with the energy scale inferred in Ref.~\cite{Hentrich} from $\kappa_{xx}(T)$.

As discussed in Refs. \cite{Ponomaryov1,Ponomaryov2,Winter}, the sharp ESR mode $C$ is increasingly dominant at large $B$ but the broad background excitation continuum persists. From this viewpoint, the observed saturation of ${\cal C}_{\rm obs}$ to 1 above 9 T in Fig. \ref{figOmega}d is physically appealing. As the dominant excitations become increasingly magnon-like, the topological Hall current, expressed as ${\cal C}_{\rm obs}$, approaches 1. For $B<$ 7 T (zig-zag state) the vanishing of $\kappa_{xy}$ (aside from the negative dip near 6.3 T) suggests that $\Omega$ is very small.

The physical picture that emerges is that the planar $\kappa_{xy}$ derives from spin excitations that live at the relatively high energy scale $\omega_1\sim$ 1 meV (11.6 K). A large Berry curvature drives these excitations as an edge-mode thermal current, which results in an Onsager-type thermal Hall current whose magnitude corresponds to a Chern number of 1 at fields above 9 T. At all $B$, the bosonic character leads to strong suppression of $\kappa_{xy}$ to near zero below 3 K. Because this strong $T$ dependence precludes a fermionic description, a critical review of reports of a half-quantized $\kappa_{xy}/T$ in $\alpha$-RuCl$_3$ seems warranted. Finally, we note that $\omega_1$ is not a bulk ``spin gap''. The sharp modes coexist with a broad continuum corresponding to spin excitations that extend well below $\omega_1$. 
We emphasize that the magnons are only well-defined above $\sim$10 T~\cite{Balz}. As we lower $B$ into the spin-liquid state, the excitations are increasingly less magnon-like, as seen in the deviations from the fit to Eq. \ref{KH} as well as the steep decrease of $C_{\rm obs}$ (Fig. \ref{figOmega}c). The deviations provide clues to the excitations in the spin-liquid state. In the low-$T$ limit, the edge-mode engendered $\kappa_{xy}$ vanishes altogether while novel bulk features emerge to define the oscillations observed in $\kappa_{xx}$~\cite{Czajka,Sodemann}.

\vspace{1cm}
\centerline{* ~~~ * ~~~  *}

\newpage
\vspace{1cm}\noindent
$^{\S}$Corresponding author email: npo@princeton.edu\\
$^\dagger$ Present address of MH: \\

\vspace{5mm}\noindent
{\bf Acknowledgements} \\
We acknowledge useful discussions with Patrick Lee and Inti Sodemann. The research was supported by the U.S. Department of Energy (award DE-SC0017863). NPO was supported by the Gordon and Betty Moore Foundation's EPiQS initiative through grant GBMF4539.

\vspace{3mm}
\noindent
{\bf Author contributions}\\
PC performed the measurements and analysed the data together with NPO, who proposed the experiment. MH developed many of the experimental techniques and provided guidance on interpreting the results. NQ measured the crystal dimensions. AB, PLK, and SEN provided guidance on prior results. The sample was grown at Oak Ridge National Laboratory by PLK, JY and DGM. The manuscript was written by PC and NPO with input from all authors.

\vspace{3mm}
\noindent
{\bf Additional Information}\\
Supplementary information is available in the online version of the paper.
Correspondence and requests for materials should be addressed to NPO.

\vspace{3mm}
\noindent
{\bf Competing financial interests}\\
The authors declare no competing financial interests.

\end{document}